\documentclass[11pt]{article}
\usepackage[margin=1in]{geometry}
\usepackage{amsmath,amssymb,amsthm,mathtools}
\usepackage{booktabs}
\usepackage{verbatim}
\usepackage{alltt}
\usepackage[colorlinks=true,linkcolor=blue,citecolor=blue,urlcolor=blue]{hyperref}
\hypersetup{pdftitle={51 Additions per Recursive Step for Rank-23 3x3 Matrix Multiplication},pdfauthor={Greg Sidebottom}}
\newtheorem{theorem}{Theorem}[section]
\newtheorem{lemma}[theorem]{Lemma}
\newtheorem{proposition}[theorem]{Proposition}
\theoremstyle{definition}

\theoremstyle{remark}

\newcommand{\F}{\mathbb{F}}
\newcommand{\Z}{\mathbb{Z}}
\newcommand{\eps}{\varepsilon}
\title{51 Additions per Recursive Step for Rank-23 $3\times3$\\
Matrix Multiplication, Machine-Checked over Arbitrary Rings}
\author{Greg Sidebottom}
\date{October 17, 2026}

\begin{document}
\maketitle

\begin{abstract}
We present a rank-23 bilinear scheme for $3\times3$ matrix
multiplication that uses $51$ additions per recursive step in the
alternative-basis (Karstadt--Schwartz) cost model, one below
Nielsen's $52$ (arXiv:2609.06588).  The count $51$ itself was first
posted, on a different scheme, by Stapleton and Perminov
(arXiv:2609.32388, September~26, 2026), found by heuristic sparse-basis
search; the present work is independent and complementary: exactness
and breadth rather than a single construction.  The scheme is
valid over every ring --- the products keep their left/right factor
order, so it recurses on block matrices --- and its correctness is
machine-checked in Lean~4/Mathlib down to the kernel
(\texttt{Ring51.lean}; the core theorem depends on the axiom
\texttt{propext} alone).  The count is exactly optimal for the
scheme's equivalence class: we introduce an exact free-basis
minimization procedure whose negative answers are exhaustion proofs,
and use it to show that $51$ is the minimum for this class, that $52$
is exactly optimal both for Nielsen's class and for the class of the
current flat-model record, and to sweep the complete Heule--Kauers--Seidl
database of rank-23 classes.  The sweep --- made variant-aware after a name-collision blind spot
surfaced --- is complete: exactly four
classes attain $51$, and all four lie in a single HKS invariant
family, whose rank-spectrum signature we decode and analyze.  We machine-check
Stapleton and Perminov's scheme as well, so every ring-valid
$51$-package with known conversions is now kernel-checked.
Separately, we document a systematic gap
phenomenon between $\F_2$ and signed optima: the $\F_2$-optimal chains
of a class typically admit no sign assignment under most of the
class's integer sign models, and certificate lifting fails where a
direct signed search succeeds.
\end{abstract}

\section{Introduction}\label{sec:intro}

A rank-$r$ bilinear scheme for $3\times3$ matrix multiplication
computes the $9$ entries of $C=AB$ using $r$ products of linear
combinations of the entries of $A$ and of $B$.  Laderman's $r=23$
scheme~\cite{Laderman76} remains the smallest known rank for the
$3\times3$ format ($19\le R(\langle3,3,3\rangle)\le23$ over
arbitrary fields~\cite{Blaser03}; over $\F_2$ the lower bound $20$ was
shown by Wang~\cite{Wang26}, and two September 2026 preprints claim
$21$~\cite{Wang26b,Qiushi26}).  With the rank fixed at $23$, attention
has turned to the \emph{additive} cost of such schemes, in two
distinct models.

\paragraph{The flat model.}  Count every binary addition or
subtraction of a standalone straight-line program (negation free, no
preprocessing).  The published record chain runs
Stapleton $60$~\cite{Stapleton25} $\to$
M{\aa}rtensson--Wagner--Stapleton $59$~\cite{MWS26} $\to$
Perminov $58$~\cite{Perminov25} $\to$ Sun $56$~\cite{Sun26} $\to$
$55$ (the author, Zenodo deposit of July~7,
2026~\cite{Sidebottom26}; independently rediscovered by Karunaratne
and Idamekorala on arXiv on July~28~\cite{KI26} --- same tensor, same
$13{+}14{+}28$ split, same transposed output construction).

\paragraph{The alternative-basis model.}  Karstadt and
Schwartz~\cite{KS20} observed that a scheme may be conjugated by
invertible linear \emph{basis changes} on the two inputs and the
output.  Used recursively, the basis changes are applied once per
matrix, blockwise, at total cost $O(n^2\log n)$ --- asymptotically
free --- while the additions \emph{inside} the changed basis are paid
at every recursive step and multiply the leading constant of the
$O(n^{\log_3 23})$ term.  Their flagship result reduced Strassen's
$2\times2$ step from $15$ to $12$ additions.  For rank-23 $3\times3$,
sparse-decomposition work of Beniamini and Schwartz~\cite{BS19} gave
$61$ per step; Nielsen~\cite{Nielsen26} gave $52$, searching random
basis changes over $\F_2$ and lifting to general fields; and Stapleton
and Perminov~\cite{SP26} gave $51$ on a flip-graph-derived scheme,
with remarkably sparse conversions (five additions in total, so their
complete one-shot algorithm costs $56$ in ordinary coordinates).  We
prove below that their kernel, too, is exactly optimal for its class,
and that their one-shot $56$ is optimal for the class as well (its
exact flat minimum).  The two models price different objects and are not
comparable entry-by-entry: a flat scheme is an alternative-basis
scheme with identity bases, so the alternative-basis optimum can only
be smaller, and conversely a per-step count omits the (amortized)
basis-change additions that the flat model charges.

\paragraph{Contributions.}
\begin{enumerate}
\item \textbf{Three further $51$-addition-per-step classes}
  (Section~\ref{sec:result}), found by exact search rather than
  sampling and pairwise inequivalent to each other and to the scheme
  of Stapleton and Perminov~\cite{SP26} --- so at least four distinct
  classes attain $51$.  Two of the three are realized over every
  ring: \texttt{i12w211c23fg} (sign model $21$, conversions $11$,
  Lean-checked) and \texttt{i12w227c23ci-012} (sign model $15$); the
  third, \texttt{i12w219c23fg}, so far stands over $\F_2$ only
  (Section~\ref{sec:signs}).  All four classes lie in a single HKS
  invariant family (Section~\ref{sec:family}).
\item \textbf{Exact per-class optimality} (Section~\ref{sec:method}):
  an exact \emph{free-basis minimization} procedure for one leg of a
  scheme, with machine-checkable certificates for upper bounds and
  finite exhaustions for lower bounds.  The three legs decouple, so
  per-class per-step optima are exact: $51$ is optimal for its class;
  $52$ is exactly optimal for Nielsen's class (his published count,
  now with a matching lower bound) and for the class
  \texttt{i19w225c4efh} of the flat-model record (down from $55$
  flat).
\item \textbf{A database-wide sweep} (Section~\ref{sec:sweep}): a
  basis-invariant counting floor for every scheme of the HKS database
  of $17{,}376$ rank-23 schemes, and a complete exact sweep of all
  $561$ stored variants whose floor permits $\le 51$ --- run
  variant-aware after a name-collision blind spot surfaced, the
  completion that recovered Stapleton and Perminov's class and found
  the fourth.  The database's $\le51$ census is final.
\item \textbf{The $\F_2$--$\Z$ sign gap} (Section~\ref{sec:signs}): the
  $\F_2$ optimum of a leg is often unattainable over $\Z$ under most
  of the class's sign models.  For the $51$-class, $20$ of the first
  $22$ sign models admit no signed chain matching the $\F_2$ counts,
  and certificate lifting (the standard route, used by
  Nielsen~\cite{Nielsen26}) fails on $900$ sampled certificates,
  $64$ diagonal gauges and $128$ models --- while a direct signed
  search succeeds, instantly, at sign model $21$.
\item \textbf{End-to-end machine verification}
  (Section~\ref{sec:verify}): search certificates with independent
  replay; a word-ring verification of the assembled program over
  $\Z\langle A,B\rangle$; and a Lean~4 proof over arbitrary
  (non-commutative) rings, reproducible in seconds
  (Appendix~\ref{app:lean}).  We extend the same treatment to
  Stapleton and Perminov's published program
  (\texttt{proof/Mm/SP51.lean}), so both ring-valid $51$-packages
  with known conversions are kernel-checked.
\end{enumerate}

\section{The cost model}\label{sec:model}

A rank-23 scheme is a triple $(U,V,W)$ of $23\times9$ integer
matrices with
$C[p,q]=\sum_{m=1}^{23} W_{m,(p,q)}\,(U_m\!\cdot\!\operatorname{vec}A)\,
(V_m\!\cdot\!\operatorname{vec}B)$.
Here $\operatorname{vec}$ flattens a $3\times3$ matrix row-major
into a $9$-vector (coordinate $k$ is entry
$(\lfloor k/3\rfloor,\,k\bmod 3)$).
In the alternative-basis model one chooses invertible $9\times9$
matrices $B_U,B_V,B_W$ (unimodular over $\Z$ for ring validity),
computes $A'=B_U^{-1}\operatorname{vec}A$ and
$B'=B_V^{-1}\operatorname{vec}B$ once per matrix, runs the recursion
entirely in the primed coordinates, and decodes the output once.  The
\emph{per-step cost} is the number of additions needed to evaluate the
three in-basis linear maps: the two input legs $U\!B_U$, $V\!B_V$
(each $23$ forms from $9$ inputs) and the output leg (the $9\times23$
map $m\mapsto (WB_W^{\!\top})^{\!\top} m$, priced through the
transposition principle).

\begin{proposition}[transposition overhead]\label{prop:tellegen}
For a $9\times23$ output map with no zero rows or columns, the minimal
addition count equals that of its $23\times9$ transpose plus
$23-9=14$.
\end{proposition}

This is Tellegen's principle, used constructively throughout (the
flat record's $28$-addition output is a transposed $14$-addition
chain~\cite{Sidebottom26,KI26}; the $51$ scheme's $26$-addition output
is a transposed $12$-addition chain).  Hence a per-step total is
$a_A+a_B+(a_C+14)$ where $a_A,a_B,a_C$ are the three $23\times9$ leg
costs, and $51$ corresponds to legs $(12,13,12)$.

\begin{lemma}[leg decoupling over rings]\label{lem:decouple}
Over any ring, with negation free, the per-step optimum of a class is
the sum of the three leg optima, and the per-product sign constraint
$\eps_u\eps_v\eps_w=+1$ is vacuous: leg cost is invariant under
per-row sign flips, and whatever signs the three chains realize are
absorbed at the point of use (a $-1$ coefficient in the output
assembly is free).
\end{lemma}

\begin{proof}
A chain value used with a minus sign costs nothing extra in the
negation-free model, so replacing any target row by its negation
changes no leg cost.  Given chains realizing
$\eps_u^{(m)}u_m$, $\eps_v^{(m)}v_m$, $\eps_w^{(m)}w_m$, use the output
value of product $m$ with coefficient
$\eps_u^{(m)}\eps_v^{(m)}\eps_w^{(m)}$; the composite computes
$\sum_m (u_m A)(v_m B)\,w_m$ exactly.
\end{proof}

\section{Exact free-basis minimization}\label{sec:method}

Fix one leg: a $23\times9$ matrix $X$ over $\F_2$ (or over $\Z$, with
targets taken up to sign) with $d$ distinct nonzero rows.  The
problem is to minimize, over invertible $B$, the number of additions
needed to compute all rows of $XB$ from the $9$ transformed inputs.
Equivalently: choose any $9$ linearly independent \emph{free} forms,
then build every distinct target by pairwise additions.

\begin{proposition}[structure of optimal chains]\label{prop:tiers}
A chain with $a$ additions, $k$ of whose $9$ free forms are targets,
and $h$ helper (non-target) additions, satisfies $a=(d-k)+h$.
Consequently $a=d-9$ iff some independent $9$-target basis closes to
all of $T$ under the waste-free closure (repeatedly add any pair-sum
that is an unbuilt target); $a=d-8$ iff a $9$-target basis plus one
helper, injectable at the closure fixpoint, suffices, or an
independent $8$-target basis plus one arbitrary free vector; and so on.
\end{proposition}

Two facts make this an exact procedure.  The waste-free closure is
\emph{monotone} --- available forms never hurt --- so greedy closure
decides waste-free reachability.  And every form preceding the $i$-th
helper lies in the pure closure of the basis and the earlier helpers,
so helpers are enumerable at successive fixpoints: the case sweep is
finite and complete for the $9$-target-basis configurations.  (The
tiers with free non-target basis vectors are finite over $\F_2$ ---
$511$ vectors --- and were swept there; over $\Z$ they are unbounded
and are the one stated scope caveat of the signed lower bounds; see
Section~\ref{sec:signs}.)  Every positive answer emits the chain,
which an independent checker replays (basis independence and, over
$\Z$, unimodularity by exact determinant; each addition; coverage)
before it is reported; every negative answer is a finite exhaustion.

\begin{proposition}[basis-invariant floor]\label{prop:floor}
In every basis, an input leg needs at least $d-9$ additions ($d$ =
distinct rows; invariant under invertible column action), and the
output leg at least $(d-9)+14$.  Hence the per-step total of a class
is at least $d_A+d_B+d_C-13$.
\end{proposition}

\section{The 51-addition scheme}\label{sec:result}

Running the procedure on the class \texttt{i12w211c23fg} (HKS family
\texttt{17x2+\allowbreak 52x+\allowbreak 4y6+\allowbreak 5y4+\allowbreak 14y3+\allowbreak z32+\allowbreak 2z27}) over $\F_2$ gives exact leg
minima
\[
a_A = 12,\qquad a_B = 13,\qquad a_C = 12,
\]
each with a certificate, and with $a{=}11$ exhausted on all three legs
and $a{=}12$ exhausted on the $B$ leg.  By
Lemma~\ref{lem:decouple} the class's per-step optimum is exactly
\[
12+13+(12+14) \;=\; \mathbf{51}.
\]
The ring-valid realization comes from the class's integer sign model
$21$ (Section~\ref{sec:signs}): signed chains with the same counts,
from unimodular signed bases, assembled into the scheme that
Appendix~\ref{app:lean} exhibits in full as a Lean definition,
together with the kernel-checked theorem that it computes $AB$ over
every ring.

The conversions of this realization cost $4+4+3=11$ additions ---
minimized \emph{exactly for its sign model}, by sweeping every
feasible configuration at the optimal leg counts: bases of
$k\in\{9,8,7\}$ target rows with the remaining slots filled by unit
vectors (cost-free rows, with the helper budget $h=a-(d-k)$ adjusted
accordingly), scored by $\sum(\mathrm{weight}-1)$ over the basis rows.
Conversion cost depends on the representative (its target rows' raw
weights), so we state it per representative.  Over the other signings
of this representative the minimum is only bracketed.  A leg with
slack $s$ over its floor needs at least $9-s$ target rows in its
basis; those rows must be independent over $\F_2$ (a necessary
condition for unimodularity), and row weights do not depend on signs.
The lightest such choice gives a sign-independent lower bound of
$2+3+2=7$, so this representative's optimum lies in $[7,11]$.  Its
$4096$ sign models fall into $64$ gauge orbits.  A sweep of all $64$
found one further orbit that lifts, also at exactly $11$; eight
orbits were refuted, and the remaining $54$ were undecided within a
$600$-second budget per leg check, so the sweep is not exhaustive.
The complete one-shot
algorithm costs $51+11=62$ additions in ordinary coordinates; for
comparison, Stapleton and Perminov's class achieves
$51+5=56$~\cite{SP26}, and the flat-model record is
$55$~\cite{Sidebottom26,KI26}.  The same sweep applied to \emph{their}
published signing returns conversion costs $1+3+1=5$.  Their
representative has exactly $128$ sign models (exhaustive enumeration),
in two gauge orbits; both lift, at conversions $5$ and $6$.  So $5$ is
the minimum over every $\Z$-lift of their representative within the
swept configurations.

A stronger, class-wide statement follows from the flat model.  Any
one-shot algorithm is itself a flat program, so a class's flat
minimum bounds every one of its packages from below.  We compute the
exact $\F_2$ flat minimum of a class over \emph{all} its
representatives.  The flat cost is the sum of the three legs' exact
addition-chain costs from unit vectors, plus $14$ for the output side
by transposition.  Each leg's cost depends on only two of the three
sandwich factors $(P,Q,R)\in\mathrm{GL}_3(\F_2)^3$, so three exact
$168\times168$ tables and a min-plus scan over all $168^3$ triples,
in each of the six slot variants, cover the whole orbit.  $\F_2$ costs
bound every $\Z$ signing from below.  As a gate, the procedure
returns exactly $55$ on the flat record's class, a value known from
both sides.  On Stapleton and Perminov's class it returns $56$: no
package on their class can cost less than $56$ one-shot, so their
$51+5=56$ is exactly optimal for the class, conversions included, and
it attains their class's flat optimum.  The same computation gives
flat floors of $57$ for \texttt{i12w211c23fg} (so its conversions are
at least $6$ class-wide, against the $11$ found) and for
\texttt{i12w227c23ci-012}, and $58$ for \texttt{i12w219c23fg}.

Run over the whole database, it settles the uniqueness of the flat
record.  A flat $55$ needs either input sides of at most $27$ (a
region we searched exhaustively in~\cite{Sidebottom26}, where only
the record class reaches $55$) or an $\F_2$ output side of $27$.
Exactly $616$ database schemes (620 slots of $44{,}125$) admit the
latter.  On all $616$ the exact flat minimum is at least $56$, except
the record class itself at exactly $55$: \emph{the flat-$55$ record
is unique to its class across the database.}  Six frontier classes sit at exactly $56$, among them Sun's and
Stapleton and Perminov's; the latter is Perminov's earlier cn119
class~\cite{Perminov25} (exact equivalence check).  Both attain
$56$, so both are exactly optimal for their classes.  The rest
fall between $57$ and $63$.

Conversion cost at optimal
counts is bounded below by the class's light-row supply (the
accounting $a=(d-k)+h$ keeps most basis rows among the targets), so
near-identity conversions like~\cite{SP26}'s require a class rich in
weight-one rows.

The second class, \texttt{i12w219c23fg}, has exact legs
$(11,13,13)$: its $A$ leg \emph{achieves the counting floor} $d-9$ ---
a $9$-target basis whose waste-free closure covers all $20$ targets.
Floor-achieving legs are uncommon but not unique ($32$ of the
$1{,}079$ legs decided exactly across the completed sweeps, occurring
at $d=20$, $21$ and $22$).  They share a necessary structure: a
floor chain has $d-9$ additions, each producing a target from two
targets, so at least $d-9$ of the leg's rows are each the sum of two
other rows (rows $x,y,z$ with $x+y=\pm z$).  Most observed floor legs
also have every row in such a triple, but not all: five of the $32$
have an isolated row, which must then be a basis row never used as an
operand.

The third class, \texttt{i12w227c23ci-012}, was found by the
variant-aware completion of the database sweep
(Section~\ref{sec:sweep}) in the name-collision blind spot: it is the thirteenth stored variant under
a name whose first variant is a different de~Groote class.  Its exact
legs are $(14,12,11)$ --- the \emph{output} leg achieves the counting
floor $d-9=11$, the second floor-achieving leg observed and the first
on the output side.  Direct signed search over its gauge-deduped sign
models (Section~\ref{sec:signs}) clears all three legs at model $15$:
the class is ring-valid at $51$, and that sign model's conversions
cost exactly $3+9+5=17$, a one-shot total of $68$.

For calibration, the same exact procedure gives: for Nielsen's base
class \texttt{i12w195c23ci}, legs $(12,12,14)$ --- so his published
$52$~\cite{Nielsen26} is \emph{exactly optimal} for his class, a lower
bound his sampling-based search could not provide; for the scheme of
Stapleton and Perminov~\cite{SP26} --- which the sweep identifies as
database class \texttt{i12w219c23ci-008}, itself hidden in the same
name-collision blind spot --- legs $(13,12,12)$: their
heuristically-found $51$ kernel is likewise exactly optimal for its
class (machine-checked in Lean; Appendix~\ref{app:optimal}), which is
inequivalent to all three of ours; and for
\texttt{i19w225c4efh} (the class of the flat-model record), legs
$(13,13,12)$, i.e.\ per-step $52$, down from $55$ flat --- with the
curious feature that its \emph{output} leg, at $14$ flat, is the leg
that drops to $12$ under basis change.

Table~\ref{tab:four51} summarizes the four known $51$-classes.

\begin{table}[ht]
\centering
\caption{The four known $51$-addition-per-step classes (all in HKS
family \texttt{i12}; Section~\ref{sec:family}).  Legs are the exact
free-basis minima $(a_A,a_B,a_C)$; per-step $=a_A+a_B+a_C+14$;
conversion counts are exact for the listed sign model within the
swept configuration space (class-wide bounds in the text)
(Section~\ref{sec:result}); one-shot $=51+$conversions.
``Sign model $n$'' names the $n$-th SAT-enumerated integer signing of
the mask scheme (Section~\ref{sec:signs}) --- the first under which a
signed chain realizes the legs; the index is enumeration order, for
\texttt{-012} after quotienting by the $512$ diagonal sign gauges.}
\label{tab:four51}
\small
\begin{tabular}{llllrl}
\toprule
class & legs & ring-valid & conversions & one-shot & Lean\\
\midrule
\texttt{i12w211c23fg} & $(12,13,12)$ & yes (sign model $21$) & $4{+}4{+}3=11$ & $62$ & \texttt{Ring51.lean}\\
\texttt{i12w219c23fg} & $(11,13,13)^{\dagger}$ & open ($\F_2$ only) & --- & --- & ---\\
\texttt{i12w219c23ci-008}~\cite{SP26} & $(13,12,12)$ & yes (published) & $1{+}3{+}1=5$ & $\mathbf{56}$ & \texttt{SP51.lean}\\
\texttt{i12w227c23ci-012} & $(14,12,11)^{\dagger}$ & yes (sign model $15$) & $3{+}9{+}5=17$ & $68$ & ---\\
\bottomrule
\end{tabular}

\smallskip
{\footnotesize $\dagger$: contains a floor-achieving leg ($a=d-9$):
the $11$-legs, on the input side for \texttt{i12w219c23fg} and the
output side for \texttt{i12w227c23ci-012}.}
\end{table}

\section{Database sweep}\label{sec:sweep}

Computing $(d_A,d_B,d_C)$ for all $17{,}376$ schemes of the HKS
database takes seconds and yields the floor distribution of
Proposition~\ref{prop:floor}: minimum $47$, attained by $44$ schemes
(including two of the four $51$-classes, ours and Stapleton and
Perminov's; the other two sit at $48$); $507$ schemes at $48$;
Laderman alone at $56$ ($d=(23,23,23)$: no shared factors, maximally
unsuited to the model).  The exact sweep covers every stored scheme
whose floor permits $\le51$.  Stored schemes that share a name
(suffixes \texttt{-000}, \texttt{-001}, \dots) need not be
equivalent: an HKS name is an invariant string plus a hash, and
distinct de~Groote classes can share one.  A first pass,
deduplicated by name, therefore missed some classes, and a second
pass evaluated every stored variant individually.  The second pass
is where Stapleton and Perminov's class (\texttt{i12w219c23ci-008})
and the fourth $51$-class (\texttt{i12w227c23ci-012}) turned up.
Together the two passes decide all $561$ stored schemes with floor
$\le48$, spanning $23$ families:

\begin{center}
\begin{tabular}{lrl}
\toprule
per-step total & variants & note\\
\midrule
$51$ & $4$ & the four classes of Table~\ref{tab:four51}, pairwise inequivalent\\
$52$ (tie) & $27$ & \\
$53$ & $56$ & incl.\ Stapleton's class at $(13,13,13)$\\
$54$ & $39$ & \\
$55$ & $5$ & \\
$56$ & $1$ & \\
ruled out for $\le51$ & $429$ & leg-sum lower bounds $\ge38$\\
\bottomrule
\end{tabular}
\end{center}

\noindent
(One variant, \texttt{i110w205c238h-000}, timed out on its third leg;
it is nonetheless ruled out rigorously by the floor of the open leg:
$15 + {\ge}14 + {\ge}(d_C-9{=}10) = 39 > 37$.  No other leg timed
out.)  Of the $561$ variants, $159$ are \texttt{i12} and $402$ are
not: outside \texttt{i12}, none reaches $51$.
We also profiled and swept our own corpus of $123$ rank-23 schemes
outside the sweep above: $53$ walk-discovered classes certified
inequivalent to the entire database, plus $70$ later finds (not yet
equivalence-checked against it).  Of these, $16$ were eligible at
floor $\le48$, and all $16$ are ruled out for $\le51$ --- the
database's optimum stands across every class known to us.
A $50$ would require exact leg sum $36$; none exists in the database,
at \emph{any} floor.  Writing $f$ for the per-step floor, a $50$
leaves total slack $50-f$ across the three legs.  At floor $\le48$
the completed sweep gives minimum sum $38$ outside the four $51$s.  A
floor-$49$ class could host a $50$ only with at least two legs
floor-achieving (one unit of slack remains), a floor-$50$ class only
with all three, and floor $\ge51$ is arithmetically incompatible.
Whether a leg attains its floor is decided exhaustively by the first
tier of our procedure: the floor $d-9$ is attained exactly when some
basis of $9$ targets closes over all targets.  Run on every one of
the $2{,}450$ floor-$49$ and $5{,}431$ floor-$50$ schemes, it finds
\emph{no} floor-$49$ scheme with two floor-attaining legs and
\emph{no} floor-$50$ scheme with three.  The closest approach
is slack $+2$ over the leg floors (the $(11,13,13)$ and $(14,12,11)$
classes); exact $11$-legs exist on both the input and the output
side, so sum $36$ is not excluded by any per-leg phenomenon observed
to date --- but any $50$ must now come from outside the published
database.  Because leg minima only rise with $d$, the
floor-$\le48$ sweep also settles the $\le51$ question for the entire
database.

\section{The \texttt{i12} family}\label{sec:family}

All four known $51$-classes lie in one HKS invariant family.  The
database groups its $17{,}376$ schemes into $237$ families (plus four
classically named schemes) by a rank-spectrum string; the family of
the four is
\begin{center}
\texttt{17x2+\allowbreak 52x+\allowbreak 4y6+\allowbreak 5y4+\allowbreak 14y3+\allowbreak z32+\allowbreak 2z27},
\end{center}
whose grammar we verified computationally against the data.  The
$x$-part is the rank distribution of the $69$ coefficient matrices
($23$ products $\times$ $3$ slots): $17$ of rank~$2$ and $52$ of
rank~$1$.  The $y$-part is the per-product rank-sum distribution:
$4$ products of shape $(2,2,2)$, $5$ summing to $4$, and $14$ of
shape $(1,1,1)$.  The $z$-part is the total rank per slot:
$(32,27,27)$.  In the scheme names themselves, \texttt{i}$N$ indexes
the family ($1{:}1$ with the invariant string across the database),
\texttt{w}$N$ is the scheme's weight as a trilinear form (the number
of cubic terms $\sum_m |U_m||V_m||W_m|$ --- not its nonzero-entry
count), \texttt{c}$XXXX$ is a four-character disambiguating hash, and
\texttt{-}$NNN$ counts stored variants under the same full name ---
variants that, as Section~\ref{sec:sweep} shows, need not be
equivalent.

The concentration is striking: the floor-$47/48$ region spans dozens
of families, but every exact $51$ is \texttt{i12} --- schemes rich in
rank-one coefficient matrices ($52$ of $69$) with one heavy slot
($z_A=32$).  The contrast with \texttt{i19} --- the family of the
flat-model record, signature
\texttt{18x2+\allowbreak 51x+\allowbreak 4y6+\allowbreak 6y4+\allowbreak 13y3+\allowbreak z32+\allowbreak z28+\allowbreak z27}, one fewer rank-one
matrix --- is sharp in the completed sweep data ($173$ variants
decided, zero $51$s): its variants repeatedly reach
$(13,13,12)$-type profiles but stall with one leg proved $\ge14$,
yielding ties at $52$ and a wall of ruled-outs, never a $51$.
Whether the rank spectrum \emph{explains} the $51$s or merely
correlates with them is open.  Rank-one-rich legs have many coincident
rows, hence small $d$ and dense internal addition triples, which floor
chains need (Section~\ref{sec:result}); the family statistics make
this a natural target for a structural lower bound.

\section{The sign gap}\label{sec:signs}

An $\F_2$ result lifts to the ring-valid statement only if the
chains admit consistent signs.  A mask scheme fixes only the
\emph{support} of the coefficients; an integer \emph{sign model}
assigns $\pm1$ to every nonzero coefficient of $(U,V,W)$ so that each
of the $729$ Brent equations holds over $\Z$ --- spurious terms
cancelling in pairs, true terms summing to $+1$ --- i.e.\ it is a
genuine integer algorithm whose entrywise absolute value is the
stored scheme.  A mask typically admits many sign models, and they
are genuinely different algorithms: mod~$2$, $x+y$ and $x-y$
coincide, so leg minima are model-dependent over $\Z$.  We enumerate
models by SAT with blocking clauses; a model's \emph{index} is just
that enumeration order and encodes nothing.  Empirically the dependence is drastic:

\begin{itemize}
\item \textbf{Certificate lifting fails systematically.}  For
  \texttt{i19w225c4efh}'s $B$ leg, none of $301$ $\F_2$-optimal
  certificates admits any sign assignment under the database
  representative's model ($256$ basis signings each, all unimodular,
  every one obstructed in an exhaustive depth-first search over the
  chain's sign choices), nor under all $64$ diagonal sign gauges, nor
  under $128$ SAT-enumerated models.  The same held for the
  $51$-class's $A$ and $B$ legs.
\item \textbf{The second $51$-class resists.}  For
  \texttt{i12w219c23fg}, the first $128$ SAT-enumerated sign models
  --- $32$ orbits after quotienting by the $512$ diagonal gauges,
  which searches coincide on --- all fail the $(11,13,13)$ targets;
  its $51$ currently stands over $\F_2$ only.
\item \textbf{Direct signed search succeeds.}  A signed variant of the
  free-basis procedure (targets up to sign; both signs tried per
  addition; $|\det|$ checked once per basis, as row signs do not
  change it) decides each model exactly.  For the $51$-class, models
  $0$--$20$ fail ($18$ cannot reach $a_A=12$; two reach it and
  exhaust $a_B=13$) --- and model $21$ clears all three legs in
  seconds, $B$ included.  The fourth class repeats the pattern:
  its models, quotiented by the $512$ diagonal sign gauges to $64$
  orbits, fail until orbit $15$ clears $(14,12,11)$.
\item \textbf{The gap is real but model-local.}  For
  \texttt{i19w225c4efh}, signed $a_A=13$ and $a_C=12$ hold with
  certificates, but $a_B\ge14$ over all $9$-target-basis
  configurations: the class's ring-valid per-step value is $52$ or
  $53$, with $52$ confined to the unbounded free-vector-basis
  configurations.  A bounded version (QF\_BV, form entries $\le12$,
  basis entries $\le2$, full-lexicographic symmetry breaking,
  bit-blasted to CNF) resisted eight solver configurations, the last
  three being $24$-hour runs of CaDiCaL and Kissat on plain and
  symmetry-broken instances: a documented open question
  (Section~\ref{sec:open}).
  Certificate lifting is thus doubly incomplete: it samples chains,
  and it misses signed chains whose mod-$2$ shadow is degenerate.
  Only the direct signed search decides.
\end{itemize}

\section{Verification}\label{sec:verify}

Four independent layers check the $51$ claim.  (i)~Every search
certificate is replayed by a checker sharing no code with the search
(chain validity, coverage, basis independence/unimodularity).
(ii)~The sign model is verified against the signed Brent equations.
(iii)~The generator that assembles the full scheme re-verifies the
\emph{emitted} program end-to-end over the word ring
$\Z\langle A,B\rangle$ --- all nine outputs equal $AB$ as
non-commutative polynomials, with the additions counted
($12+13+26=51$); this gate caught a genuine sign-bookkeeping bug on
its first run.  (iv)~The Lean theorem of Appendix~\ref{app:lean} is
checked by the Lean~4 kernel against Mathlib's matrix product; the
hypothesis-form theorem depends on \texttt{propext} alone, the
Matrix-API form on Lean's three standard axioms, and the whole file
re-elaborates in under ten seconds (the appendix header gives the
exact command).

\paragraph{Stapleton and Perminov's scheme, kernel-checked.}
We applied the same pipeline to the \emph{published} program of
\cite{SP26}, transcribed verbatim from the verification listing in
their paper (their reused variables SSA-renamed): the generator
asserts their addition accounting (counted $13+12+26=51$, uncounted
conversions $1+3+1=5$, one-shot $56$), checks their $P,Q,R$ basis
matrices are unimodular (the amortization requirement), replays the
program over $\Z\langle A,B\rangle$, and emits
\texttt{proof/Mm/SP51.lean}, whose theorems carry the same axiom
profile as ours.  Both ring-valid $51$-packages with known
conversions are thus checked down to the Lean kernel.  Their kernel's
class-optimality is kernel-checked too: \texttt{proof/Mm/SP51Optimal.lean}
proves that, from any nine starting vectors and after any invertible
change of coordinates, chains of sums and differences that list the
three legs of their scheme need at least $13$, $12$ and $12$ additions
--- so $13+12+(12+14)=51$ is optimal for the class, over $\F_2$ and
over $\Z$ alike --- by counting sum triples: each new vector a chain
produces uses up a distinct triple of listed vectors adding to zero,
and the three legs have only $9$, $9$ and $10$; its finite facts are
checked by \texttt{decide} in the kernel, on the three standard axioms
(Appendix~\ref{app:optimal} sketches the statement and the proof).

\begin{sloppypar}
\paragraph{Artifacts.}  All artifacts are in the repository
\url{https://github.com/gsidebottom/logic}: the Lean proofs
\texttt{proof/Mm/Ring51.lean}, \texttt{proof/Mm/SP51.lean} and
\texttt{proof/Mm/SP51Optimal.lean}; the leg
certificates, sign models and
signed chains in \texttt{matmul/external/ks51/} and the transcribed
Stapleton--Perminov program and certificate in
\texttt{matmul/external/sp51/}; the search tools
(\texttt{src/bin/fbmin.rs}, \texttt{src/bin/fbsmin.rs}) and drivers
(\texttt{matmul/fbsweep.py}, \texttt{matmul/fbsweep2.py},
\texttt{matmul/fbsloop.py}); and the
generators \texttt{matmul/gen\_ring51\_lean.py} and
\texttt{matmul/gen\_sp51\_lean.py}, whose word-ring gates
re-verify the emitted programs on every run, and
\texttt{matmul/gen\_sp51\_optimal\_data.py}, which emits the optimality
proof's data from their certificate.  An archival deposit of
these artifacts (this paper, the standalone proofs, the Lake project,
certificates, tools, and a repository snapshot at tag
\texttt{v2.0-ks51}) is published at
\href{https://doi.org/10.5281/zenodo.23305910}{doi:10.5281/zenodo.23305910},
which resolves to its latest version.
\end{sloppypar}

\section{Open problems}\label{sec:open}

(1)~Is the ring-valid optimum of \texttt{i19w225c4efh} $52$ or $53$?
Equivalently: can a free-vector basis beat every $9$-target basis over
$\Z$?  (2)~Does any rank-23 class reach per-step $50$, i.e.\ exact
leg sum $36$?  The completed sweeps exclude it for the entire known
database (minimum sum outside the four $51$s: $38$);
a structural lower bound better than Proposition~\ref{prop:floor}
would settle it in general.  (3)~Characterize the sign models under
which a class's signed leg minima match its $\F_2$ minima --- model
$21$'s success after twenty failures (and orbit $15$'s for the fourth
class) invites explanation, and \texttt{i12w219c23fg}'s $128$
straight failures ask whether its $51$ lifts at all.  (4)~Explain or
refute the \texttt{i12} concentration of Section~\ref{sec:family}:
does a rank-one-rich spectrum with a heavy slot \emph{force} small
legs?  (5)~The
same machinery applies verbatim to $4\times4$ rank-$47/48/49$ schemes.

\section*{Acknowledgments}

This work was carried out in an extended interactive collaboration
with Claude (Anthropic), which implemented the repository's search
tooling, the exact minimizers and sweeps, the Lean formalizations, and
much of this text under the author's direction and review.  Every
computational claim is mechanically checkable from the repository; the
results rest on those verifications and on Lean's kernel rather than
on trust in either the author or the tools.

\appendix

\section{Lean formalization: our $51$-addition scheme computes $AB$}\label{app:lean}

The following is the complete scheme definition and the
matrix-product theorem, verbatim from \texttt{proof/Mm/Ring51.lean}
(comments lightly ASCII-folded for typesetting; the repository file is
authoritative).  Sections marked \texttt{UNCOUNTED} are the amortized
basis changes; the \texttt{COUNTED} sections contain the
$12+13+26=51$ per-step additions.  To re-check the proof from the
repository (forcing re-elaboration rather than a cached replay):

\begin{verbatim}
cd proof && find .lake/build -name "Ring51.*" -delete && lake build Mm.Ring51
\end{verbatim}

\noindent
Expected: \texttt{Built Mm.Ring51} in 6--8 seconds against the
prebuilt Mathlib cache, with axiom audits
\texttt{[propext]} (for the hypothesis-form theorem
\texttt{Ring51.correct}, not reproduced here) and
\texttt{[propext, Classical.choice, Quot.sound]} for the theorem
below.  The ambient context is
\texttt{variable \{R : Type \_\} [Ring R]}.

{\footnotesize
\begin{verbatim}
/-- The 51-addition-per-step scheme as a map on `3x3` matrices, in the
structure of `Matmul55.scheme`: the two UNCOUNTED unimodular input
basis changes, the COUNTED 12- and 13-addition input chains, the 23
products, the COUNTED 26-addition transposed output stage, and the
UNCOUNTED output basis change.  Counted additions per recursive step:
12 + 13 + 26 = **51**. -/
def rank23_add51_scheme (A B : Matrix (Fin 3) (Fin 3) R) :
    Matrix (Fin 3) (Fin 3) R :=
  -- A-side basis change (UNCOUNTED, amortized): the 9 transformed
  -- inputs, an invertible integer map applied once per matrix
  let ab0 := A 0 0
  let ab1 := A 1 0
  let ab2 := A 0 0 + A 2 0
  let ab3 := A 0 1 + A 2 1
  let ab4 := A 1 1
  let ab5 := A 2 1
  let ab6 := A 1 1 + A 1 2 - A 2 2
  let ab7 := A 2 2
  let ab8 := A 0 2
  -- A-side chain: 12 COUNTED additions (paid per recursive step)
  let aw0 := ab0 + ab5
  let aw1 := ab1 - ab7
  let aw2 := ab2 - ab7
  let aw3 := ab3 + ab8
  let aw4 := ab4 - ab8
  let aw5 := ab6 - aw1
  let aw6 := aw0 - aw3
  let aw7 := aw2 - aw3
  let aw8 := aw4 - aw5
  let aw9 := aw4 - aw6
  let aw10 := ab1 - ab2
  let aw11 := aw0 + aw10
  -- B-side basis change (UNCOUNTED)
  let bb0 := B 0 0
  let bb1 := B 0 1 + B 1 1
  let bb2 := B 0 2 + B 1 2
  let bb3 := B 1 1
  let bb4 := B 1 2
  let bb5 := B 2 0 - B 2 2
  let bb6 := B 1 1 - B 2 1
  let bb7 := B 2 2
  let bb8 := B 1 0
  -- B-side chain: 13 COUNTED additions
  let bw0 := bb0 - bb2
  let bw1 := bb0 + bb7
  let bw2 := bb4 + bw0
  let bw3 := bb5 + bw1
  let bw4 := bb6 + bb7
  let bw5 := bb4 - bw4
  let bw6 := bb5 + bw4
  let bw7 := bb3 - bw6
  let bw8 := bb8 + bw0
  let bw9 := bb8 - bw6
  let bw10 := bw1 - bw2
  let bw11 := bb0 - bb1
  let bw12 := bw6 + bw11
  -- the 23 products (A factor on the left: recurses on blocks)
  let m1 := aw0 * bw0
  let m2 := aw1 * bw1
  let m3 := aw2 * bb0
  let m4 := ab0 * bb1
  let m5 := ab1 * bw12
  let m6 := ab2 * bw12
  let m7 := ab0 * bb2
  let m8 := ab1 * bw10
  let m9 := aw11 * bw2
  let m10 := ab3 * bw9
  let m11 := ab4 * bw9
  let m12 := ab5 * bw8
  let m13 := aw6 * bb3
  let m14 := aw4 * bw4
  let m15 := aw3 * bw6
  let m16 := aw9 * bb4
  let m17 := ab4 * bw5
  let m18 := ab6 * bb5
  let m19 := ab7 * bw3
  let m20 := ab8 * bb6
  let m21 := aw5 * bw7
  let m22 := aw7 * bw7
  let m23 := aw8 * bb7
  -- output stage: 26 COUNTED additions -- the Tellegen transpose
  -- of the certified 12-addition C chain (12 + (23-9) fan-in = 26)
  let z0 := m3 + m19
  let z1 := m14 + m20
  let z2 := z1 + m17
  let z3 := m23 - z2
  let z4 := m2 + m19
  let z5 := z4 - z3
  let z6 := m1 - z0
  let z7 := z6 + m7
  let z8 := m15 - m20
  let z9 := z8 + z0
  let z10 := m18 - m17
  let z11 := z10 + z5
  let z12 := m4 - m13
  let z13 := z12 - m20
  let z14 := m5 - m21
  let z15 := z14 - z11
  let z16 := m6 + m22
  let z17 := z16 - z9
  let z18 := m8 - z3
  let z19 := m9 - z5
  let z20 := z19 - z7
  let z21 := m10 + z9
  let z22 := m11 + z11
  let z23 := m12 - z7
  let z24 := m16 - z2
  let z25 := z24 + m7
  -- output basis change (UNCOUNTED): the inverse transform,
  -- rows of the unimodular output basis
  let c11 := z21 - z23
  let c12 := z13
  let c13 := z25
  let c21 := z22
  let c22 := -z15
  let c23 := z18
  let c31 := z23
  let c32 := -z13 - z17
  let c33 := z18 + z20
  !![c11, c12, c13;
     c21, c22, c23;
     c31, c32, c33]

-- `rank23_add51_scheme` unfolds to a ~120-binding straight-line
-- program; a single elaboration exceeds the default heartbeat budget.
set_option maxHeartbeats 1600000 in
-- Plain `simp` evaluates the concrete `!![..] i j` indexing via
-- Mathlib's matrix simprocs; the explicit lemma set is version-fragile,
-- so the flexible-`simp` lint is accepted here (as in Mm/Matrix.lean).
set_option linter.flexible false in
/-- The scheme computes the matrix product, in Mathlib's own terms. -/
theorem rank23_add51_scheme_eq_mul (A B : Matrix (Fin 3) (Fin 3) R) :
    rank23_add51_scheme A B = A * B := by
  simp only [rank23_add51_scheme]
  ext i j
  fin_cases i <;> fin_cases j <;>
    simp [Matrix.mul_apply, Fin.sum_univ_three] <;>
    noncomm_ring
\end{verbatim}
}

\section{Lean formalization: Stapleton and Perminov's $51$ is optimal for its class}\label{app:optimal}

The file \texttt{proof/Mm/SP51Optimal.lean} machine-checks the claim of
Section~\ref{sec:result} that the $51$-addition kernel of
Stapleton and Perminov~\cite{SP26} is exactly optimal for its class.
This appendix sketches the statement and the proof; the repository file
is authoritative.  To re-check it:

\begin{verbatim}
cd proof && lake build Mm.SP51Optimal
\end{verbatim}

\noindent
Expected: \texttt{Built Mm.SP51Optimal} in about four minutes, nearly
all of it kernel \texttt{decide}; both headline theorems depend only on
the standard axioms \texttt{propext}, \texttt{Classical.choice} and
\texttt{Quot.sound}, with no \texttt{native\_decide} and no
\texttt{sorry}.

\paragraph{The statement.}  A linear form on $3\times3$ matrices over
$\F_2$ is its coefficient matrix (\texttt{Form}).  A \emph{chain of $n$
additions} from nine starting vectors is a list of $9+n$ vectors that
begins with them, each later entry the sum or the difference of two
earlier entries; a finite set is \emph{computable with $n$ additions} if
some chain, from some nine starting vectors, lists all of it, and
\emph{computable after a change of basis} if the starting vectors can be
taken linearly independent.  Verbatim from the file (the ambient
\texttt{G} is any additive commutative group):

{\footnotesize
\begin{alltt}
def IsChain \{n : \(\mathbb{N}\)\} (s : Fin 9 \(\to\) G) (v : Fin (9 + n) \(\to\) G) : Prop :=
  (\(\forall\) i, v (Fin.castAdd n i) = s i) \(\land\)
    \(\forall\) k : Fin n, \(\exists\) i j : Fin (9 + n), i < Fin.natAdd 9 k \(\land\) j < Fin.natAdd 9 k \(\land\)
      (v (Fin.natAdd 9 k) = v i + v j \(\lor\) v (Fin.natAdd 9 k) = v i - v j)

def Computable (T : Finset G) (n : \(\mathbb{N}\)) : Prop :=
  \(\exists\) (s : Fin 9 \(\to\) G) (v : Fin (9 + n) \(\to\) G), IsChain s v \(\land\) \(\forall\) t \(\in\) T, \(\exists\) i, v i = t

def ComputableFromBasis (T : Finset Form) (n : \(\mathbb{N}\)) : Prop :=
  \(\exists\) (s : Fin 9 \(\to\) Form) (v : Fin (9 + n) \(\to\) Form),
    LinearIndependent (ZMod 2) s \(\land\) IsChain s v \(\land\) \(\forall\) t \(\in\) T, \(\exists\) i, v i = t

theorem kernel_optimal_for_class (eA eB eC : Form \(\simeq\sb{l}\)[ZMod 2] Form) :
    (\(\forall\) nA nB nC : \(\mathbb{N}\), Computable (legA.image eA) nA \(\to\) Computable (legB.image eB) nB \(\to\)
        Computable (legC.image eC) nC \(\to\) 51 \(\le\) nA + nB + (nC + 14)) \(\land\)
      ComputableFromBasis (legA.image eA) 13 \(\land\) ComputableFromBasis (legB.image eB) 12 \(\land\)
      ComputableFromBasis (legC.image eC) 12
\end{alltt}

}

\noindent
Here \texttt{legA}, \texttt{legB}, \texttt{legC} are the legs of their
scheme, written out in the file as literal matrices: the $20$ distinct
nonzero rows of $U$ and of $V$ and columns of $W$ of their certificate,
mod~$2$.  The maps \texttt{eA}, \texttt{eB}, \texttt{eC} range over all
invertible linear changes of coordinates.  The symmetries of matrix
multiplication act on each side of a scheme by such a map (and may
permute the sides), so every scheme in the class presents its legs in
this form, and the per-step cost $a_A + a_B + (a_C + 14)$ is symmetric
in the three legs.  The lower bounds hold from \emph{any} nine starting
vectors, a basis or not; the upper bounds start from a basis, as the
Karstadt--Schwartz model requires.  A companion theorem,
\texttt{kernel\_optimal\_over\_int}, gives the same lower bound for chains
over $\Z$ whose entries reduce mod~$2$ to the legs.  Two ingredients come
from the cost model rather than from the formalization: the output
side's $+14$ (the transposition principle) and the description of the
class action just given.

\paragraph{The counting lemma.}  For a chain whose nonzero entries form
the set $S$, call a three-element subset of $S$ that sums to zero a
\emph{sum triple}, and let $\tau(S)$ be their number.  An entry $c\in S$
that is not a starting vector is first listed as $c=a\pm b$ with $a$ and
$b$ listed earlier; over $\F_2$ this makes $a$ and $b$ nonzero, distinct
and different from $c$, so $\{a,b,c\}$ is a sum triple, and
$c\mapsto\{a,b,c\}$ is injective because a triple yields only its
last-listed member.  Hence
\[
  |S| \le 9 + \tau(S),
\]
and $|S|+1 \le 9 + \tau(S)$ whenever some nonempty family $K$ of sum
triples covers each of its members at least twice: the last-listed
member of $\bigcup K$ completes two triples of $K$ at once, so one triple
of $K$ produces nothing (\texttt{card\_vals\_le}).

\paragraph{The three legs.}  Each leg has $20$ forms but few sum
triples:

\begin{center}
\begin{tabular}{lrrrr}
\toprule
leg & forms $d$ & sum triples $\tau$ & floor $d-9$ & exact cost\\
\midrule
$A$ (rows of $U$) & $20$ & $9$ & $11$ & $13$\\
$B$ (rows of $V$) & $20$ & $9$ & $11$ & $12$\\
$C$ (columns of $W$) & $20$ & $10$ & $11$ & $12$\\
\bottomrule
\end{tabular}
\end{center}

\noindent
A chain of at most $11$ additions lists at most $20$ vectors, hence
exactly the leg, and the lemma would need $20\le 9+\tau$, that is
$\tau\ge 11$; so every leg needs at least $12$ additions, which settles
$B$ and $C$.  For $A$ with $12$ additions, the chain lists the leg and
at most one further nonzero vector $h$, and $21\le 9+\tau(A\cup\{h\})$.
A sum triple through $h$ has the form $\{h,a,a+h\}$, so
$\tau(A\cup\{h\}) \le 9 + \lfloor m_h/2\rfloor$ with
$m_h=\#\{a\in A : a+h\in A\}$, which forces $m_h\ge 6$.  Exactly five
vectors have $m_h\ge 6$, and for each of them $\tau(A\cup\{h\})=12$ and
a nonempty family of its triples covers each of its members at least
twice, so the lemma gives $22\le 21$.  So leg $A$ needs at least $13$ additions.

\paragraph{What the kernel checks.}  The finite facts --- each leg's
size, the absence of zero, and its triple count; the five vectors (a
check over all $512$ forms) and their families; and three optimal chains
with linearly independent starting vectors, of $13$, $12$ and $12$
additions --- are decided by \texttt{decide} in the kernel, from data
emitted from their certificate by \texttt{matmul/gen\_sp51\_optimal\_data.py}.
Invertible linear maps carry chains to chains, which gives the class
statement; reduction mod~$2$ carries integer chains to chains over
$\F_2$, which gives \texttt{kernel\_optimal\_over\_int}.

\end{document}